\documentclass{article}
\usepackage[english]{babel}

\usepackage{amsmath,amssymb,tabularx,bm,array,booktabs,authblk}
\usepackage{xcolor,slashed,epsfig,cite}
\usepackage{verbatim,slashed,ulem,geometry,setspace,algorithm}
\usepackage{algorithmic,mathrsfs,braket,multirow,dcolumn,subfigure}
\usepackage{graphicx}
\usepackage{threeparttable}
\graphicspath{{./Figures/}}
\usepackage[colorlinks=true,citecolor=blue,urlcolor=cyan,filecolor=magenta,linkcolor=red,frenchlinks=true,bookmarks]{hyperref}
\allowdisplaybreaks[4]
\renewcommand\sout{\bgroup \color{red}  \ULdepth=-.5ex \ULset}

\begin{document}

\title{Form factors of decuplet baryons in a covariant quark-diquark approach}

\newcommand{\email}[1]{\thanks{\href{mailto:#1}{#1}}}

\author[a,b]{JiaQi Wang\email{jqwang@ihep.ac.cn}}
\author[a,b]{Dongyan Fu\email{fudongyan@ihep.ac.cn}}
\author[a,b]{Yubing Dong\email{dongyb@ihep.ac.cn}}

\affil[a]{Institute of High Energy Physics, Chinese Academy of Sciences, 
Beijing 100049, China}
\affil[b]{School of Physical Sciences, University of Chinese Academy 
of Sciences, Beijing 101408, China}

\maketitle
\begin{abstract}

The electromagnetic and gravitational form factors of decuplet baryons are systematically studied with a covariant quark-diquark approach. The model parameters are firstly discussed and determined through comparison with the lattice calculation results integrally. Then, the electromagnetic properties of the systems including electromagnetic radii, magnetic moments, and electric-quadrupole moments are calculated. The obtained results are in agreement with experimental measurements and the results of other models. Finally, the gravitational form factors and the mechanical properties of the decuplet baryons, such as mass radii, energy densities, and spin distributions, are also calculated and discussed. 

\end{abstract}

\section{\MakeUppercase{INTRODUCTION}}

\quad\, Form factors (FFs) provide a wealth of information for comprehending the inner structures of 
particles. Electromagnetic form factors (EMFFs) could provide the electromagnetic properties of a system, such as its charge radius, magnetic moment, and even higher-order moments. Meanwhile, gravitational form factors (GFFs), which are derived from the matrix element of the symmetric energy-momentum tensor~\cite{Polyakov_2018}, could give the mechanical properties such as the mass and angular momentum distributions.
    
\begin{figure}[htbp]
    \centering
    \includegraphics[width=0.3\linewidth]{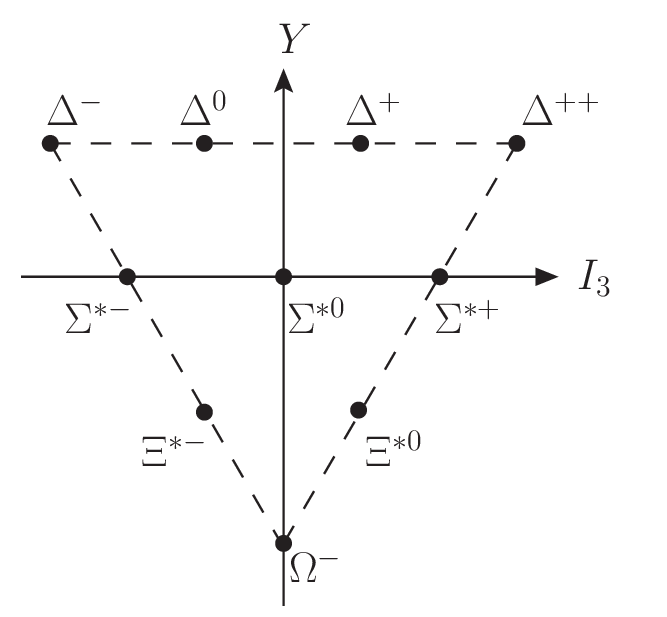}
    \caption{\small{Members of the decuplet baryons.}}
    \label{decuplet}
\end{figure}

The spin-3/2 particle is the main research object in this work.
The most fundamental spin-3/2 particles, including $\Delta(1232)$, $\Sigma(1385)$, $\Xi(1530)$, and $\Omega(1672)$, are known as the decuplet baryons with SU(3) symmetry, and it is important to investigate them systematically. The composition of the decuplets is illustrated in Fig.~\ref{decuplet}. The $\Delta$ resonance, as the lowest excited state of the nucleon, has been considered as a typical target in the research of spin-3/2 particles. Unfortunately, due to short lifetime~\cite{ParticleDataGroup:2022pth} of the $\Delta$ isobar, directly measuring its EMFFs in the experiment remains a challenge. In the decuplets, $\Sigma^*$ and $\Xi^*$ have the similar short lifetime, while $\Omega^-$ has a longer lifetime with $c\tau=2.461$ cm~\cite{ParticleDataGroup:2022pth}. 
Fortunately, the transition processes are expected to yield information on accessing the electromagnetic properties of $\Delta$ and other decuplet baryons~\cite{Blanpied:1997zz,CLAS:2001cbm,Sparveris:2013ena}. Additionally, the magnetic moments of $\Delta^{++}$ and $\Delta^{+}$ have been measured through $\pi^+ p \rightarrow \pi^+ p \gamma$~\cite{LopezCastro:2000cv} and $\gamma p \rightarrow \pi^0 p \gamma'$~\cite{Kotulla:2002cg} processes. Since $\Omega^-$ has a longer lifetime, there are more opportunities to directly probe its structure, and then its time-like form factors and effective 
form factors have been measured by CLEO~\cite{Dobbs:2014ifa} and BESIII~\cite{PhysRevD.107.052003} through the process of $e^+e^-\rightarrow \Omega^- \overline{\Omega}^+$. Furthermore, we expect that the coming experimental facilities may provide us more useful data to understand the electromagnetic structures of the decuplets. For example, BESIII and possible future super $J/\psi$ factory SCTF are expected to bring the secondary beam of $\Omega^-$ through $\psi(2S) \rightarrow \Omega^-\overline{\Omega}^+$ process~\cite{PhysRevLett.127.012003} and JLab (Jefferson Lab) is planning to measure the electromagnetic properties of $\Sigma^*$ and $\Xi^*$ in future experiments~\cite{PhysRevC.76.025208,CLAS:2018kvn}.

Although some experiment facilities are working on the EMFFs of the decuplet baryons, it is still hard to measure their GFFs directly due to the negligible gravitational interaction. However, the GFFs can be extracted from the generalized parton distributions~\cite{Diehl:2003ny,Fu:2022bpf,PhysRevD.101.096008,Fu:2023dea} and the generalized distribution amplitudes~\cite{PhysRevD.97.014020}. With respect to the nucleon, generalized parton distributions are expected to be measured from deeply virtual Compton scattering~\cite{Ji:1998pc} at some facilities including JLab~\cite{Burkert:2018bqq}, the future EIC (Electron Ion Collider)~\cite{doi:10.1142/11684}, and the EicC (Electron-Ion Collider in China)~\cite{Chen:2018J+}.

There have been plenty of theoretical works about the FFs of hadrons for decades, including those with targets of spin-0 ~\cite{Hohler:1974eq,Maris:2000sk,PhysRevD.97.014020,PhysRevD.78.094011,PhysRevD.101.096008}, spin-1/2~\cite{Bincer:1959tz,Keiner:1996at,Meyer:1994cn,Ma_Qing_Schmidt_2002,Kim_Schweitzer_Yakhshiev_2012,Goeke_Grabis_2007,Marcucci:2015rca}, and spin-1~\cite{PhysRevD.99.094035,Sun:2017gtz,Dong:2013rk,PhysRevD.100.036003}. The electromagnetic properties of the decuplets with spin-3/2 have also been studied with various approaches, such as lattice QCD (LQCD)~\cite{Alexandrou_2009,Alexandrou:2010jv,Boinepalli:2009sq,PhysRevD.79.051502,Leinweber:1992hy,Lee:2005ds}, the Skyrme model~\cite{Oh:1995hn}, the chiral perturbation theories ($\chi$PT)~\cite{Geng:2009ys,Li:2016ezv}, the quark models~\cite{Krivoruchenko:1991pm,Berger:2004yi,Schlumpf:1993rm}, the QCD sum rules~\cite{Aliev:2009pd,Lee:1997jk,Azizi:2009egn}, the chiral constituent quark model ($\chi$CQM)~\cite{Wagner:2000ii}, the 1/$N_c$ expansion~\cite{Flores-Mendieta:2015wir,Buchmann:2002et}, the general QCD parameterization method (GPM)~\cite{Buchmann:2002xq,Buchmann:2008zza}, the chiral quark soliton model ($\chi$QSM)~\cite{Kim_Kim_2019}, and some others.

Although some calculations have been carried out for the electromagnetic properties of the decuplet baryons in their literatures, systematical studies of their GFFs are still lacking. In this paper, we present a systematical investigation of the decuplets and simultaneously calculate their EMFFs and GFFs with a relativistic and covariant quark-diquark approach, which has been employed for two typical baryons, $\Delta$ resonance~\cite{PhysRevD.105.096002} and $\Omega^-$~\cite{Fu:2023ijy}, in our previous works. Recall that the determinations of model parameters employed in Refs.~\cite{PhysRevD.105.096002,Fu:2023ijy} are based on fitting to the LQCD results of $\Delta$ (the $u$ and $d$ quark system) and $\Omega^-$ (the $s$ quark system), respectively. In order to give a systematical description of all the decuplet baryons built from $u$, $d$, and $s$ quarks, a new set of model parameters is re-determined. Then, the EMFFs and GFFs of the systems are calculated.

This paper is organized as follows. Section~\ref{sectionapproach} briefly shows the definitions of the EMFFs and GFFs of a spin-3/2 particle and introduces our covariant quark-diquark approach simply. In Sec.~\ref{sectionresults}, the model parameters used in this calculation for the decuplet baryons are discussed and determined. With the parameters used, the obtained electromagnetic properties of the systems, including electromagnetic multipole moments and radii, are explicitly listed and we give comparisons of our results with other model calculations. Similarly, the GFFs are calculated and the mechanical properties including the mass radii, energy distributions, and angular momentum distributions in the coordinate space are discussed. Finally, Sec.~\ref{summary} is devoted to a brief summary and some discussion.

\section{FORM FACTORS AND QUARK-DIQUARK APPROACH}\label{sectionapproach}

\subsection{Electromagnetic form factors}

\quad\, For a spin-3/2 particle, the matrix element of the electromagnetic current can be parameterized as~\cite{Cotogno_2020}
\begin{equation}
    \label{ecur}
    \begin{split}
    	\left\langle p^\prime,\lambda^\prime \left| \hat{J}_a^{\mu}
    	\left( 0 \right)
    \right| p,\lambda\right\rangle=&-\bar{u}_{\alpha'} 
    \left( p',\lambda' \right)
    \biggl[ \frac{P^\mu}{M} \left( g^{\alpha' \alpha } F_{1,0}^{V,a} 
    \left( t \right)
    -\frac{q ^{\alpha'} q ^\alpha}{2M^2} F_{1,1}^{V,a} 
    \left( t \right)\right)\\
    &+\frac{i \sigma^{\mu q}}{2M} \left( g^{\alpha' \alpha} F_{2,0}^{V,a} 
    \left( t \right)
    -\frac{q ^{\alpha'} q ^\alpha}{2M^2}F_{2,1}^{V,a} 
    \left( t \right)\right) \biggr]
    u_\alpha \left( p,\lambda \right),
    \end{split}
    \end{equation}
where $i \sigma^{\mu q}=i \sigma^{\mu\rho}q_\rho$, $M$ stands for the baryon mass and $u_\alpha \left( p,\lambda\right)$ 
is the Rarita-Schwinger spinor with normalization as $\bar{u}_{\sigma'}(p) u_\sigma(p)=-2 M \delta_{\sigma' \sigma}$. 
The kinematical variables introduced in Eq.~\eqref{ecur} are defined as $P^\mu=\left(p^\mu+p'^\mu\right)/2$, 
$q^\mu=p'^\mu-p^\mu$, and $t=-q^2$, where $p$ ($p'$) is the initial (final) momentum. The index $a$ in $F_{i,j}^{V,a}$ 
runs from the quark to the gluon and the total form factor is the sum of them. In this work, we only consider the constituent quark contribution.
    
In the Breit frame, the average of the baryon momenta and the momentum transfer are defined as $P=(E,0)$ and $q=(0,\bm{q})$, where $E$ is the energy carried by the baryon. Then, the EMFFs of a spin-3/2 particle can be expressed in terms of $F_{i,j}^V$~\cite{Nozawa:1990gt}
\begin{subequations}
    \label{EMFFeq}
    \begin{align}
        G_{E0}\left( t \right)=&\left( 1+\frac{2}{3}\tau \right) 
        [F_{2,0}^V(t) + (1+\tau)(F_{1,0}^V(t)-F_{2,0}^V(t))] 
        +\frac{2}{3} \tau (1+\tau) [F_{2,1}^V(t) + 
        (1+\tau)(F_{1,1}^V(t)-F_{2,1}^V(t))],
    \\
        G_{E2}\left( t \right)=& [F_{2,0}^V(t) + 
        (1+\tau)(F_{1,0}^V(t)-F_{2,0}^V(t))] + (1+\tau) [F_{2,1}^V(t) + 
        (1+\tau)(F_{1,1}^V(t)-F_{2,1}^V(t))],
    \\
        G_{M1}\left( t \right)=&\left(1+\frac{4}{5}\tau\right) F_{2,0}^V 
        \left( t \right)
        +\frac{4}{5} \tau  (\tau +1) F_{2,1}^V \left( t \right),
    \\
        G_{M3}\left( t \right)=& F_{2,0}^V \left( t \right)
        + (\tau +1) F_{2,1}^V \left( t \right),
    \end{align}
    \end{subequations} 
where $\tau=-t/(4 M^2)$ with $t<0$. In Eq.~\eqref{EMFFeq}, $G_{E0}$, $G_{E2}$, $G_{M1}$, and $G_{M3}$ respectively represent the electric-monopole, electric-quadrupole, magnetic-dipole, and magnetic-octupole form factors. When the squared momentum transfer $t$ goes to 0, the electric charge $Q_e$, magnetic moment $\mu$, electric-quadrupole moment $\mathcal{Q}$, and magnetic-octupole moment $\mathcal{O}$ can be obtained through~\cite{RAMALHO2009355}
\begin{equation}
    \begin{split}
    	Q_e=G_{E0}(0), &\quad \mu=\frac{e}{2M} G_{M1}(0),\\
    	\mathcal{Q}=\frac{e}{M^2} G_{E2}(0), 
    	&\quad \mathcal{O}=\frac{e}{2M^3} G_{M3}(0).
    \end{split}
    \end{equation}
Moreover, the electric charge and magnetic radii are defined from their corresponding form factors as~\cite{Leinweber:1992hy}
    \begin{equation}
    	{\langle r^2\rangle}_{E0}=\left.\frac{6}{G_{E0}(0)} 
    	\frac{d}{dt}G_{E0}(t)\right|_{t=0}, 
    	\quad {\langle r^2\rangle}_{M1}=\left.
    	\frac{6}{G_{M1}(0)} \frac{d}{dt}G_{M1}(t)\right|_{t=0} \footnote{For the neutral baryon, the radii are defined as~\cite{Li:2016ezv}
\begin{equation}
	{\langle r^2\rangle}_{E0}=\left.6 \frac{d}{dt}G_{E0}(t)\right|_{t=0}, 
	\quad {\langle r^2\rangle}_{M1}=\left.6 \frac{d}{dt}G_{M1}(t)\right|_{t=0}.
	\nonumber
\end{equation}}.
    \end{equation}

\subsection{Gravitational form factors}
\quad\, The GFFs can be calculated from the matrix element of the energy-momentum tensor $\hat{T}^{\mu \nu}$ as~\cite{Cotogno_2020}
    \begin{equation}
    \label{gcur}
    \begin{split}
        &\left\langle p^\prime,\lambda^\prime \left| 
        \hat{T}^{\mu \nu}_a(0)\right| p,\lambda\right\rangle  \\
        =&-\bar{u}_{\alpha ^\prime}\left(p^\prime,\lambda^\prime\right) 
        \bigg [\frac{P^\mu P^\nu}{M}\left(g^{\alpha'\alpha} 
        F_{1,0}^{T,a} (t)-\frac{q ^{\alpha' } 
        q ^{\alpha}}{2 M^2}F_{1,1}^{T,a} (t) \right)
        + \frac{ \left({q }^\mu {q }^\nu- {g}^{\mu \nu}q^2\right)}{4M}    
        \left({g}^{\alpha'\alpha}F_{2,0}^{T,a} (t)
        -\frac{{q }^{\alpha'}{q}^{\alpha}}{2 M^2}F_{2,1}^{T,a} (t)\right)\\
        &+ M g^{\mu  \nu} \left(g^{\alpha'\alpha}F_{3,0}^{T,a} (t)
        -\frac{ q^{\alpha'} q^{\alpha}}{2M^2}F_{3,1}^{T,a} (t)\right)
        + \frac{i {P}^{ \{ \mu } \sigma ^{\nu \} \rho}q_\rho}{2M} 
        \left(g^{\alpha' \alpha}F_{4,0}^{T,a} (t) 
        -\frac{q ^{\alpha'} q ^{\alpha}}{2 M^2}F_{4,1}^{T,a}(t)\right) \\
        &- \frac{1}{M} \left({q }^{\{ \mu} g^{\nu \} 
        \{ \alpha '} {q }^{\alpha \}}-2 q^{\alpha' }q^{\alpha} g^{\mu \nu }
    - g^{\alpha' \{ \mu } g^{\nu \} \alpha } q^2 \right) F_{5,0}^{T,a} (t)    
    + M g^{\alpha ' \{ \mu } g^{\nu \} \alpha}F_{6,0}^{T,a} (t) 
    \bigg]u_\alpha \left(p,\lambda\right),
    \end{split}
    \end{equation}
where the convention $a^{\{\mu}b^{\nu\}}=a^\mu b^\nu+a^\nu b^\mu$ is used. Notice that $F_{3,0}^T$, $F_{3,1}^T$ and $F_{6,0}^T$ are the non-conserving terms which will vanish when considering the contribution from the gluon and we simply ignore them.

Analogous to the EMFFs, the gravitational multipole form factors (GMFFs), including the energy-monopole (-quadrupole) form factors ${\varepsilon}_{0(2)}(t)$, the angular momentum-dipole (-octupole) form factors $\mathcal{J}_{1(3)}(t)$, and the form factors $D_{0,2,3}(t)$, can be expressed as the linear combination of the GFFs, $F_{i,j}^T(t)$. The detailed definitions of the GMFFs have been explicitly given in Ref.~\cite{Kim:2020lrs}, and thus we do not repeat them to avoid verbosity. Moreover, the mass radius of a baryon is obtained through the energy-monopole form factor as
    \begin{equation}
    \label{MassRadius}
        \langle r^2 \rangle_M = \left.\frac{6}{\varepsilon_0(0)}
        \frac{d}{dt}  \varepsilon_0(t) \right|_{t=0}.
    \end{equation}

The energy, angular momentum, and mechanical force densities of the baryons in the coordinate space ($r$-space) can be derived through Fourier transformation into the corresponding form factors. The energy-monopole and energy-quadrupole densities are defined as~\cite{Kim:2020lrs}
    \begin{equation}
    \label{epsi}
        \mathcal{E}_0(r) = M \widetilde{{\varepsilon}}_0(r), 
        \qquad\mathcal{E}_2(r)=-\frac{1}{M}r\frac{d}{dr}\frac{1}{r}
        \frac{d}{dr}\widetilde{{\varepsilon}}_2(r),
    \end{equation}
with 
    \begin{eqnarray}
    \label{Fourie}
        \widetilde{{\varepsilon}}_{0,2}(r)=
    \int\frac{d^3 q}{(2\pi)^3}e^{-i\bm{q}\cdot\bm{r}}{\varepsilon}_{0,2}(t),
    \end{eqnarray}
being the densities in $r$-space. The angular momentum density can be expressed as
    \begin{equation}\label{rhor}
        \rho_J(r)=-\frac{1}{3}r\frac{d}{d r}
        \int\frac{d^3 q}{(2\pi)^3}e^{-i\bm{q\cdot r}}\mathcal{J}_1(t).
    \end{equation}

According to Ref.~\cite{Polyakov_2018}, it is argued that the densities of the corresponding pressure and shear force in the classical medium physics are derived from the form factors correlated with the ``D-term" as
\begin{equation}
    \label{force}
    \begin{split}
        p_0 (r) & = \frac{1}{6 M} \frac{1}{r^2} \frac{d}{d r} r^2 
        \frac{d}{d r}\tilde{D}_0 (r),\\
        s_0 (r) & = - \frac{1}{4 M} r \frac{d}{d r} \frac{1}{r} 
        \frac{d}{d r}\tilde{D}_0 (r),
    \end{split}
    \end{equation}
where
\begin{equation}
    \tilde{D}_0 (r) = \int \frac{d^3 q}{(2 \pi)^3} e^{- i \bm{q} \cdot \bm{r}} D_0 (t).
\end{equation}
The higher-order pressures and shear forces are omitted here and explicitly listed in Ref.~\cite{Kim:2020lrs}.

\subsection{Quark-diquark approach}

\quad\, We know that the decuplet baryons are composed of three quarks and have the spin of $3/2$. In our quark-diquark approach, we treat the baryon as a bound state of a spin-1/2 quark and a spin-1 (axial-vector) diquark. The SU(6) spin-flavor wave functions of the decuplets are listed in~\ref{appendix}~\cite{Lichtenberg_Tassie_Keleman_1968}.
According to the wave functions, the total matrix element can be expressed as the sum of the quark 
and diquark contributions,
\begin{equation}
        \left\langle p^\prime,\lambda^\prime \left| \hat{J}^{\mu}(0) 
        \right| p,\lambda\right\rangle 
        =\left\langle p^\prime,\lambda^\prime\left| 
        \hat{J}^{\mu}_{q}(0) \right|p,\lambda\right\rangle 
        +\left\langle p^\prime,\lambda^\prime \left| 
        \hat{J}^{\mu}_D(0)\right|p,\lambda\right\rangle.
\end{equation}

\begin{figure}[htbp]
    	\centering
    	\subfigure[]{\includegraphics[scale=0.65]{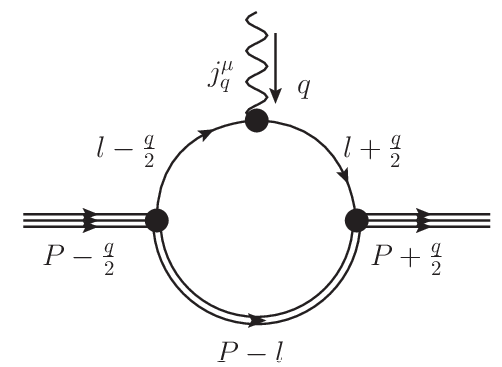}\label{jq}}
        \subfigure[]{\includegraphics[scale=0.65]{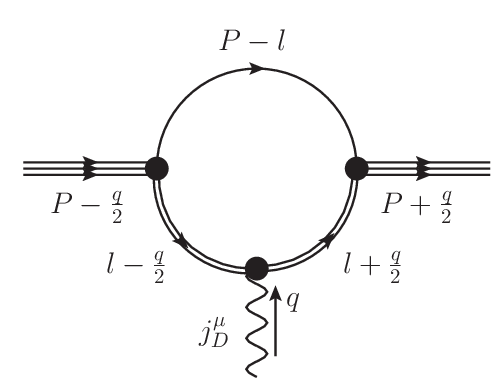}\label{jd}}
        \subfigure[]{\includegraphics[scale=0.65]{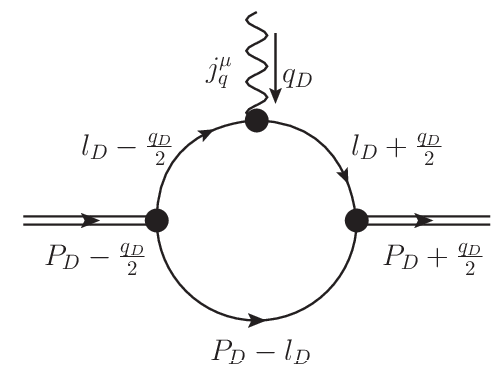}\label{jd0}}
    	\caption{\small{Feynman diagrams for the electromagnetic matrix elements 
    	contributed by the quark (a) and the diquark (b). And (c) gives the internal 
    	structure of the diquark in this process.}}
    	\label{f-emff}
\end{figure} 
Figure~\ref{f-emff} gives the Feynman diagrams for the electromagnetic interaction. One can write down the contribution of quark 
according to the Fig.~\ref{jq}
    \begin{equation}
    \label{quarkj}
    \begin{split}
        \left\langle p^\prime,\lambda^\prime \left| \hat{J}^{\mu}_{q}(0)
        \right| p,\lambda\right\rangle = & - Q^q_e e 
        \bar{u}_{\alpha'}(p',\lambda') {\left( -i \mathcal{C}^2 \right)} \\
        ~~~&\times \int \frac{d^4 l}{(2 
        \pi)^4}\frac{1}{\mathfrak{D}}\Gamma^{\alpha' \beta'} \left( 
        \slashed{l}+\frac{\slashed{q}}{2}+m_q \right) g_{\beta' 
        \beta} \gamma^\mu \left( \slashed{l}-\frac{\slashed{q}}{2} +m_q 
        \right)\Gamma^{\alpha \beta} u_{\alpha}(p,\lambda),
    \end{split}
    \end{equation}
where $Q_e^q$ is the electric charge carried by the quark, $\mathcal{C}$ is a normalization constant to ensure the calculated result $G_{E0}^q(0)=Q_e^q$, and $\Gamma^{\alpha \beta}$ is the vertex of the baryon with its quark 
and diquark constituents. Specially note that we neglect the $k^\mu k^\nu/m_D^2$ term in the propagator of the 
diquark ($1^+$) to avoid divergence of the integral~\cite{PhysRevD.79.094013}. According to Ref.~\cite{Scadron_1968}, the Lorentz structure of the vertex is
\begin{equation}
    \label{vertexfunction}        
    \Gamma^{\alpha\beta}=g^{\alpha\beta}+c_2\gamma^\beta\Lambda^\alpha
    +c_3\Lambda^\beta\Lambda^\alpha,
\end{equation}
where $\Lambda$ is the relative momentum between the quark and the diquark. The parameters of couplings $c_2$, $c_3$ in Eq.~\eqref{vertexfunction} can be 
determined by fitting to the lattice data~\cite{Alexandrou_2009,Alexandrou:2010jv} and we assume that they are independent on the baryon mass.
$\mathfrak{D}$ in Eq.~\eqref{quarkj} contains the denominators of the propagators and a special scalar function 
$\zeta$ attached to the vertex to ensure that the quark and the diquark can form a bound state. Here we simply choose the 
function~\cite{PhysRevD.80.054021}
\begin{equation}
    \label{vertexfunction2}
        \zeta (p_1,p_2)=\frac{\mathcal{C}}{\left[ p_1^2 -m_R^2
        +i \epsilon\right]\left[ p_2^2 -m_R^2+i \epsilon\right]},
\end{equation}
with $m_R$ as a cutoff parameter which is positively correlated with the baryon mass. The total $\mathfrak{D}$ is thus written as
\begin{equation}
    \label{Dfunc}
    \begin{split}
        \mathfrak{D}=&\left[\left(l-P\right)^2-m_R^2+i \epsilon\right]^2 
        \biggl[\left( l-\frac{q}{2} \right) ^2 -m_R^2+i \epsilon\biggr]
        \biggl[\left( l+\frac{q}{2} \right)^2-m_R^2+i\epsilon\biggr]\\
        &\times \biggl[\left(l+ \frac{q}{2} \right)^2-m_q^2
        +i \epsilon\biggr]\biggl[\left(l- \frac{q}{2} \right)^2 - m_q^2+i 
        \epsilon\biggr]\left[\left(l-P\right)^2-m_D^2+i \epsilon\right].
    \end{split}
\end{equation}

Similarly, the diquark contribution~\ref{jd} can be expressed as
\begin{equation}
    \label{Diquarkj}
        \left\langle p^\prime,\lambda^\prime \left| \hat{J}^{\mu}_D(0)
        \right| p,\lambda\right\rangle= - Q^D_{e} e 
        \bar{u}_{\alpha'}(p',\lambda') {i \mathcal{C}^2} 
\int \frac{d^4l}{(2\pi)^4}\frac{1}{\mathfrak{D}'}\Gamma^{\alpha'}_{~\beta'} 
        \left( \slashed{P}-\slashed{l}+m_q \right)j_D^{\mu,\beta' \beta} 
        \Gamma^{~\alpha}_{\beta} u_{\alpha}(p,\lambda),
\end{equation}
where 
\begin{equation}
    \label{Dfunc2}
    \begin{split}
        \mathfrak{D}'=&\left[\left(l-P\right)^2-m_R^2+i 
        \epsilon\right]^2 \biggl[\left( l-\frac{q}{2} \right) ^2 -m_R^2+i 
        \epsilon\biggr]\biggl
        [\left(l+\frac{q}{2} \right) ^2 -m_R^2+i \epsilon\biggr] \\
        &\times \biggl[\left(l+ \frac{q}{2} \right)^2-m_D^2+i \epsilon\biggr]
        \biggl[\left(l- \frac{q}{2} \right)^2 - m_D^2
        +i \epsilon\biggr]\left[\left(l-P\right)^2-m_q^2+i \epsilon\right].
    \end{split}
\end{equation}
$j_D^{\mu,\beta' \beta}$ in the above equation stands for the effective electromagnetic current of the diquark. Considering a diquark 
composed of quarks $q_a$ and $q_b$, the electromagnetic current can be derived from
\begin{equation}
    \label{Diquarkj2}
        \sum_{i=a,b} \left\langle p_D^\prime,\lambda_D^\prime 
        \left| \hat{J}^{\mu}_{q_i}(0)\right| 
        p_D,\lambda_D\right\rangle=-\epsilon^*_{\beta^\prime}\left(p_D^\prime,
        \lambda_D^\prime\right) j_D^{\mu,\beta'\beta} \epsilon_\beta 
        \left( p_D,\lambda_D \right),
\end{equation}
where $\epsilon_\beta(p_D,\lambda_D)$ represents the spin-1 diquark field and the kinematical variables are defined as 
$P_D^\mu=\left(p_D^\mu+p_D'^\mu\right)/2$, $q_D^\mu=p_D'^\mu-p_D^\mu=q^\mu$, and $q_D^2=-t_D=-t$. 

Assuming that the diquark is almost on shell, we can write down the matrix element
\begin{equation}
    \label{quarkj2}
    \begin{split}
        \left\langle p^\prime,\lambda^\prime \left| \hat{J}^{\mu}_{q_i}(0)
        \right| p,\lambda\right\rangle = & - Q^{q_i}_{e} e 
        \epsilon_{\beta'}^*(p'_D,\lambda'_D) 
        {\left( -i \mathcal{C}_D^2 \right)} \\
        &\times \int \frac{d^4 
        l_D}{(2\pi)^4}\frac{1}{\mathfrak{D}_D}\gamma^{\beta'}\left( 
        \slashed{l}_D+\frac{\slashed{q}_D}{2}+m_q\right)\gamma^\mu\gamma^{\beta}
        \left(\slashed{l}_D-\frac{\slashed{q}_D}{2}+m_q 
        \right)\epsilon_{\beta}(p_D,\lambda_D),
    \end{split}
\end{equation}    
where the quark-diquark vertex $\gamma^\beta$ is borrowed from Ref.~\cite{Meyer_1994}, $\mathcal{C}_D$ is the 
normalization constant similar with $\mathcal{C}$, and $\mathfrak{D}_D$ is defined as
\begin{equation}
    \label{Dfunc3}
    \begin{split}
        \mathfrak{D}_D=&\left[\left(l_D-P_D\right)^2-m_R^2+i \epsilon\right]^2 
        \biggl[\left( l_D-\frac{q_D}{2} \right) ^2 -m_R^2
        +i \epsilon\biggr]\biggl[\left( l_D+\frac{q_D}{2} \right) ^2 
        -m_R^2+i \epsilon\biggr] \\
        &\times \biggl[\left(l_D+ \frac{q_D}{2} \right)^2-m_q^2
        +i \epsilon\biggr]
        \biggl[\left(l_D- \frac{q_D}{2} \right)^2 - m_q^2
        +i \epsilon\biggr]\left[\left(l_D-P_D\right)^2-m_q^2+i \epsilon\right].
    \end{split}
\end{equation}
Finally, the effective electromagnetic current $j_D^{\mu,\beta'\beta}$ can be written as
\begin{equation}
\label{spin1ff}
    j_D^{\mu,\beta'\beta}=\left[g^{\beta'\beta}F_{D,1}^V(t)
    -\frac{q^{\beta'}q^\beta}{2m_D^2}F_{D,2}^V(t)\right]
    \left(p'_D+p_D\right)^\mu-(q^{\beta'}g^{\mu\beta}
    -q^\beta g^{\mu\beta'})F_{D,3}^V(t),
\end{equation}
where $F_{D,1 \, (2,3)}^V(t)$ are the three form factors of the spin-1 diquark. 
    
In terms of the GFFs and according to the quark Lagrangian
\begin{equation}
    \label{QuarkLag}
        \mathcal{L}=\frac{i}{2}\overline{\psi}_q\gamma^\mu{\mathop{\partial}
        \limits^{\leftrightarrow}} _\mu\psi_q+m_q\overline{\psi}_q\psi_q,
\end{equation}
with ${\mathop{\partial}\limits^{\leftrightarrow}}_\mu
={\mathop{\partial}\limits^{\rightarrow}}_\mu
-{\mathop{\partial}\limits^{\leftarrow}}_\mu$, 
we have the symmetric energy-momentum tensor of the quark as
\begin{equation}
    \label{QuarkEMT}
    	T^{\mu\nu}=\frac{i}{4}\overline{\psi}_q\gamma^
    	\mu{\mathop{\partial^\nu}\limits^{\leftrightarrow}}
    	\psi_q+\frac{i}{4}\overline{\psi}_q\gamma^\nu{\mathop{\partial^\mu}
    	\limits^{\leftrightarrow}}\psi_q.
\end{equation} 
Therefore, the GFFs contributed by the quark and the diquark can be calculated by replacing $\gamma^\mu$ 
with $\gamma^\mu l^\nu+\gamma^\nu l^\mu$ in Eqs.~\eqref{quarkj} and~\eqref{quarkj2}. Our work on $\Delta(1232)$~\cite{PhysRevD.105.096002} gives the calculation process in detail.
    
\section{NUMERICAL RESULTS}\label{sectionresults}

\subsection{Parameter determination} \label{paradtm} 

\quad\, By using the on-shell identities in Ref.~\cite{Cotogno_2020}, we can extract the form factors from Eqs.~\eqref{ecur} 
and~\eqref{gcur}. Before doing the calculation of the loop integrals numerically, it is necessary to input the model 
parameters including the baryon mass $M$, quark mass $m_q$, diquark mass $m_D$, and the cutoff parameter $m_R$ introduced 
in Eq.~\eqref{vertexfunction2}. Moreover, the couplings $c_2$, $c_3$ in the quark-diquark vertex \eqref{vertexfunction}
are also needed to be determined. It should be mentioned that in our previous studies on $\Delta$ isobar (the $u$ and $d$ quark system)~\cite{PhysRevD.105.096002} and $\Omega^-$~\cite{Fu:2023ijy} (the $s$ quark system), 
we chose two sets of parameters separately. Here, since we aim to give a systematical description of 
all the decuplet baryons, the parameters are re-determined. We simply keep the parameters $c_2$, $c_3$, and $m_R$ in 
Ref.~\cite{Fu:2023ijy} for $\Omega$ hyperon (the $s$ quark system) and re-determine the parameters associated to the light-flavor, 
like $m_{u}$, $m_{ud}$, and $m_{us}$, since the mass of $\Delta$ is defined as the average 
between $\Delta$ and nucleon instead of its physical mass in Ref.~\cite{PhysRevD.105.096002}.

In this work, all the decuplet baryon masses $M$ is chosen from Ref.~\cite{ParticleDataGroup:2022pth}. To ensure that the quark and the diquark 
are in bound states, the input masses of quark and diquark need to satisfy the relation $M<m_q+m_D$ and $m_D<m_{q_a}+m_{q_b}$. 
Since $m_R$ is positively correlated with the baryon mass and has little effect on the results~\cite{PhysRevD.105.096002,Fu:2023ijy}, we simply borrow $m_R=2.2$ GeV from our previous work about the heaviest 
baryon $\Omega^-$~\cite{Fu:2023ijy}.

\begin{figure}[htbp]
    	\centering
    	\subfigure[]{\includegraphics[width=0.32\linewidth]{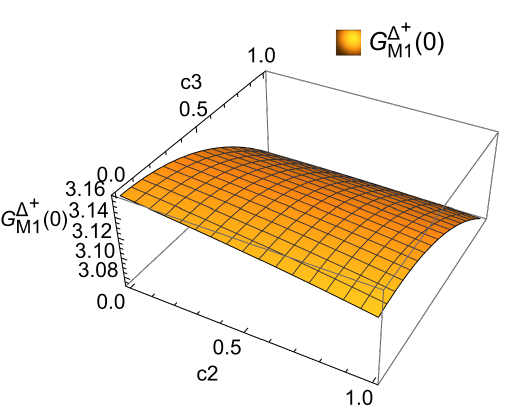}
    	\label{c2c3m1}}
        \subfigure[]{\includegraphics[width=0.32\linewidth]{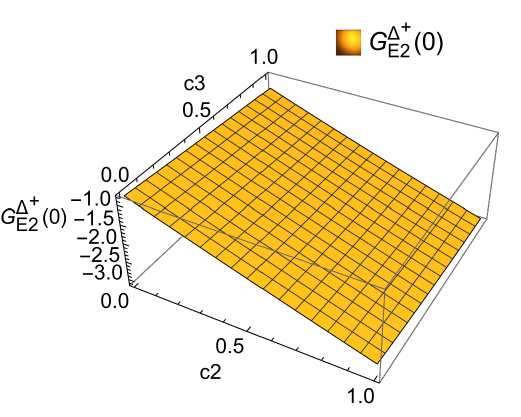}
        \label{c2c3e2}}
        \subfigure[]{\includegraphics[width=0.32\linewidth]{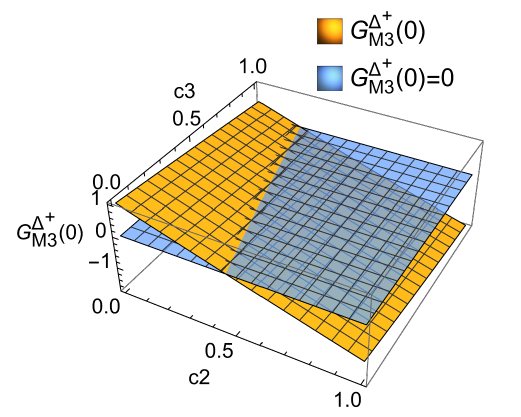}
        \label{c2c3m3}}
    	\caption{\small{$G_{M1}^{\Delta^+}(0)$, $G_{E2}^{\Delta^+}(0)$, and $G_{M3}^{\Delta^+}(0)$ as the parameters $c_2$ and $c_3$ (in units of $\text{GeV}^{-1}$ and $\text{GeV}^{-2}$ respectively) change.}}
    	\label{c2c3}
\end{figure}

As shown in Fig.~\ref{c2c3m1}, $c_2$, $c_3$ have little impact on the electric-monopole and magnetic-dipole form factors. When $c_2$ and $c_3$ (in units of $\text{GeV}^{-1}$ and $\text{GeV}^{-2}$, respectively) run from 0 to 1, the value of $G_{M1}^{\Delta^+}(0)$ only changes about 3\%. However, the higher-order multipoles, 
especially the magnetic-octupole form factor $G_{M3}(t)$, are sensitive to the values of $c_2$ and $c_3$. According to Fig.~\ref{c2c3m3}, $G_{M3}^{\Delta^+}(0)$ even changes its sign as the two parameters increase. Here we keep the same parameters from our previous work about $\Omega^-$~\cite{Fu:2023ijy}, $c_2=0.306$ GeV$^{-1}$ and $c_3=0.056$ GeV$^{-2}$, which are obtained through fitting to the LQCD data on the electric-monopole, electric-quadrupole and magnetic-dipole form factors. Since the experimental and empirical LQCD results of $G_{M3}(t)$ are still lacking, our $c_2$, $c_3$ are only roughly determined.

Finally, we get one set of parameters to describe the EMFFs and the GFFs of all the decuplet baryons simultaneously in Table~\ref{para},
where $m_{q_1 q_2}$ stands for the mass of the diquark composed of $q_1$ and $q_2$, and we assume that $m_d=m_u$, $m_{us}=m_{ds}$, and $m_{uu}=m_{ud}=m_{dd}$.
    
\begin{table}[htbp]
        \renewcommand\arraystretch{1.3}
		\centering
		\begin{tabular}{p{2cm}<{\centering} p{2cm}<{\centering} 
		p{2cm}<{\centering} p{2cm}<{\centering} p{2cm}<{\centering} 
		p{2cm}<{\centering}}
     		\toprule
     		\toprule
       		$M_\Delta$/GeV & $M_{\Sigma^*}$/GeV & $M_{\Xi^*}$/GeV & 
       		$M_\Omega$/GeV & $m_u$/GeV & $m_s$/GeV\\
        	\midrule
        	1.232 & 1.385 & 1.530 & 1.672 & 0.43 & 0.6\\ 
            \toprule
     		\toprule
            $m_{ud}$/GeV & $m_{us}$/GeV & $m_{ss}$/GeV & $m_R$/GeV 
            & $c_2$/GeV$^{-1}$ & $c_3$/GeV$^{-2}$\\
            \midrule
            0.82 & 0.99 & 1.15 & 2.2 & 0.306 & 0.056\\ 
        	\bottomrule
     	\end{tabular}
     	\caption{\small{Parameters used in this work}} 
		\label{para}  
\end{table} 
	
\subsection{EMFFs numerical results}\label{EMFFsnumericalresults}   

\begin{figure}[htbp]
	    \centering
		\includegraphics[width=0.47\linewidth]{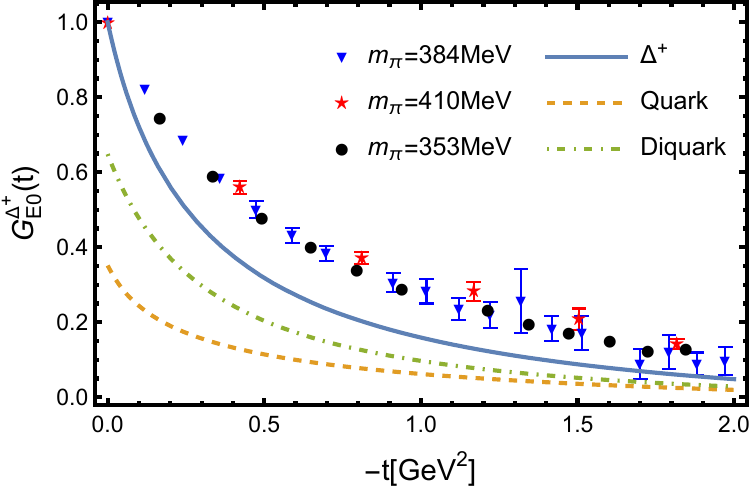}
    	\includegraphics[width=0.48\linewidth]{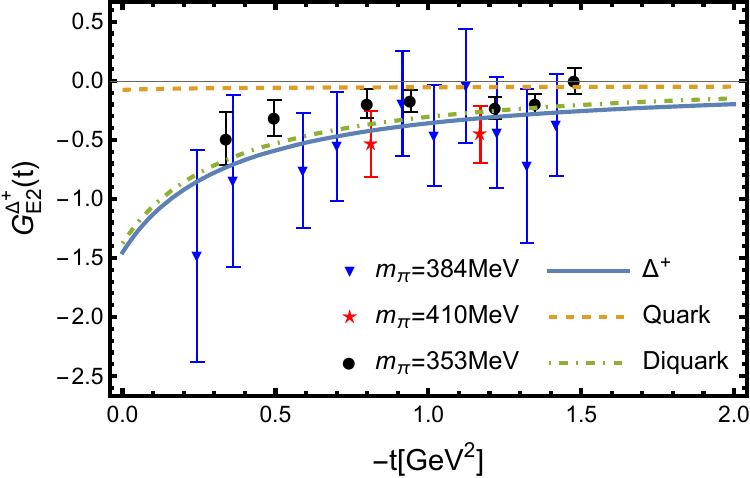}
		\includegraphics[width=0.47\linewidth]{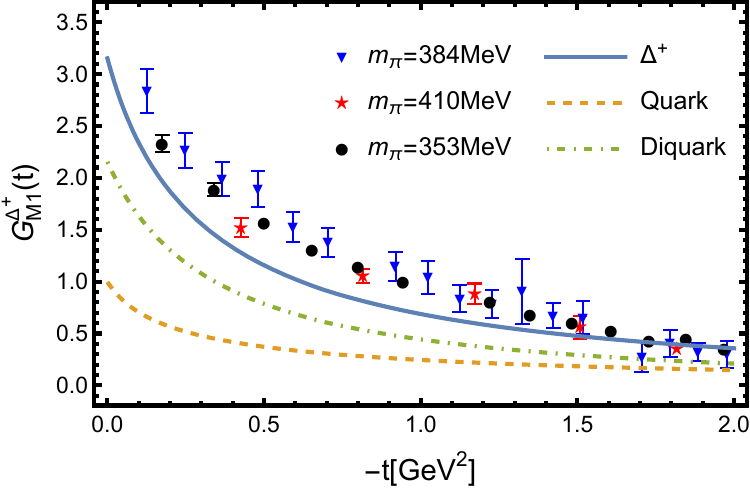}
	    \includegraphics[width=0.48\linewidth]{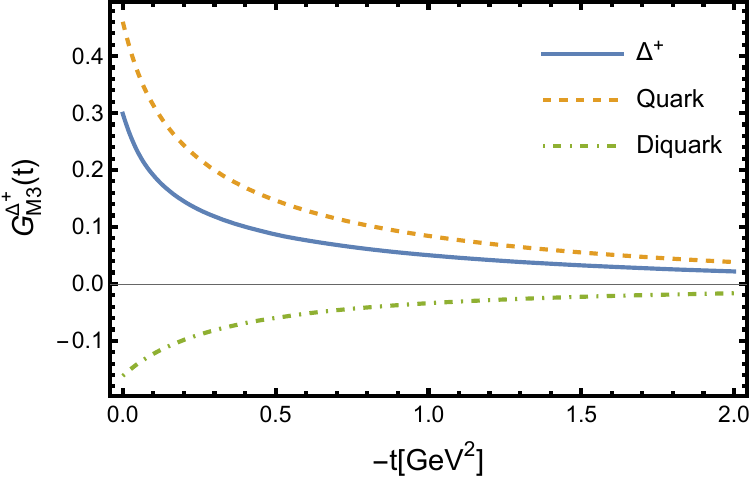}
    	\caption{\small{EMFFs of $\Delta^+$, comparing with the LQCD results 
    	\cite{Alexandrou_2009}. The solid, dashed, and dot-dashed curves 
    	represent the total EMFFs and those contributed by quark and diquark.}}
    	\label{d2EML}
\end{figure}

\quad\, Here we show our calculated results of the EMFFs of the decuplet baryons. In Fig.~\ref{d2EML}, our EMFFs of $\Delta^+$ qualitatively consistent with the LQCD results of Ref.~\cite{Alexandrou_2009} and also with our previous calculation in Ref.~\cite{PhysRevD.105.096002}.\footnote{Our $G_{E0}(t)$ are smaller than the lattice results. It should be mentioned that Ref.~\cite{Alexandrou_2009} gives the $\Delta$ isobar mass being about 1.5~GeV which is about 30\% overestimated.} The figures also show the quark and the diquark contributions separately.
Since $\Delta^+$ is composed of both $u(ud)$ and $d(uu)$, the value is their average according to the wave function 
in~\ref{appendix}. Figures.~\ref{sEM} and~\ref{xEM} plot the EMFFs of other different isospin states of $\Sigma^*$ and $\Xi^*$. 
For isovectors of $\Sigma^{*+}$, $\Sigma^{*0}$ and $\Sigma^{*-}$ , we employ the same normalization constant $\mathcal{C}$ to ensure 
$G_{E0}^{\Sigma^{*+}}(0)=1$. As seen in the first panel in Fig.~\ref{sEM}, $G_{E0}(0)$ of $\Sigma^{*0}$ and $\Sigma^{*-}$ are very close to 0 and $-1$, respectively, indicating that the normalization condition is nearly satisfied. 
The similar results occur for $\Xi^*$. It should be specially mentioned that our EMFFs results of $\Delta^0$ are strictly zero, however, those of $\Sigma^{*0}$ and $\Xi^{*0}$ are close to but not exactly zero due to $s$ and $u(d)$ have different masses, which slightly breaks the SU(3) symmetry. Since the form factors of $\Omega^-$ 
have been calculated with the same set of parameters and shown in our previous work~\cite{Fu:2023ijy}, we do not address them here for simplicity. 
    
Tables~\ref{cr} and~\ref{mr} list the electric charge and magnetic radii obtained from our 
work and other studies including LQCD~\cite{Alexandrou_2009,Alexandrou:2010jv,Boinepalli:2009sq}, chiral quark 
model~\cite{Wagner:2000ii}, $1/N_c$ expansion~\cite{Buchmann:2002et,Flores-Mendieta:2015wir} and so on. Compared with other works, 
our results are generally larger but qualitatively consistent with theirs. In our previous study on the $\Delta$ 
resonance~\cite{Fu:2022bpf}, we have chosen the baryon mass as $M=1.085$ GeV, which is the average of 
$\Delta(1232)$ and nucleon. Since we choose a different set of parameters for $u$ and $d$ quarks in 
this work, the charge radius of $\Delta(1232)$ here is a little bit lager than that in Ref.~\cite{Fu:2022bpf}.
It is seen that, for $\Delta^-$, $\Sigma^{*-}$, $\Xi^{*-}$, and $\Omega^-$ hyperons, the electric charge and magnetic 
radii decrease in turn. This feature may attribute to the different binding strengths of the baryons. $\Omega^-$ has the longest lifetime in the decuplets, suggesting that its binding strengths is the strongest. Consequently, the location of 
quarks inside $\Omega^-$ may be much close to the origin and leads to the smallest radius.
Similarly, for the $\Delta$, $\Sigma^*$ and $\Xi^{*}$ isobars, the lager decay width stands for the less stable structure, which leads 
to the lager radius.

The magnetic moments, compared with the ones from other theoretical and experimental works, of all the decuplets are given in Table~\ref{mm}. As seen in Table~\ref{mm}, ours are qualitatively consistent with the experiments and other studies. Table~\ref{eqm} shows the electric quadrupole moments, whose sign characterizes the deformation of the charge 
distribution. The positive value suggests that the particle has a prolate charge distribution, and on the contrary, the negative 
value stands for an oblate shape.
To sum up, we find that all the baryons with the positive charge have negative electric quadrupole moments, and the negatively charged 
baryons are on the opposite. Note that the obtained moments are the so-called spectroscopic moments, which are measured 
in the laboratory. Therefore, the shape discussed in this paper is the spectroscopic shape instead of the geometric shape 
derived from the intrinsic quadrupole moments~\cite{Buchmann:2001gj,Kumar:1972zza}.

\begin{figure}[htbp]
    	\centering
		\includegraphics[width=0.48\linewidth]{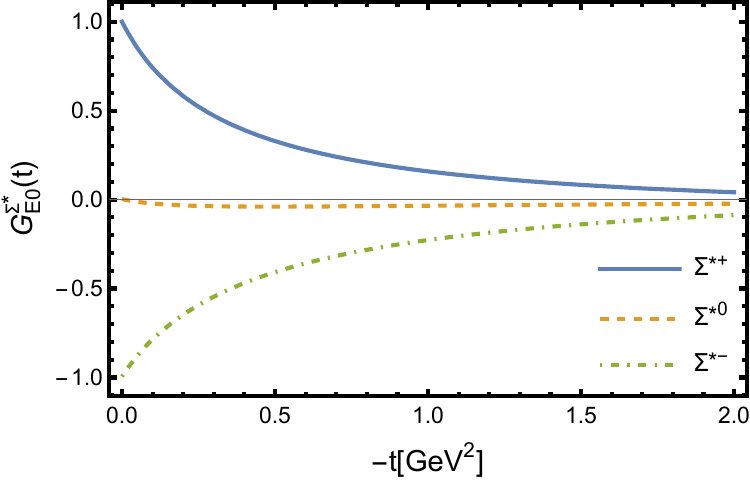}
	    \, \, \includegraphics[width=0.47\linewidth]{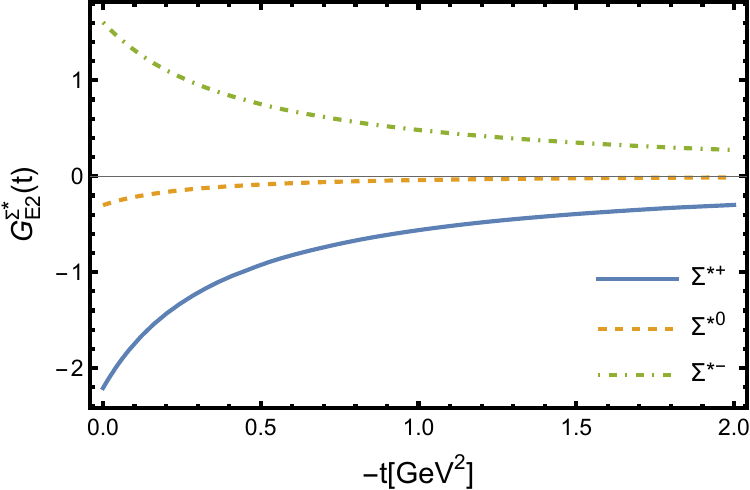}\\
		\, \includegraphics[width=0.47\linewidth]{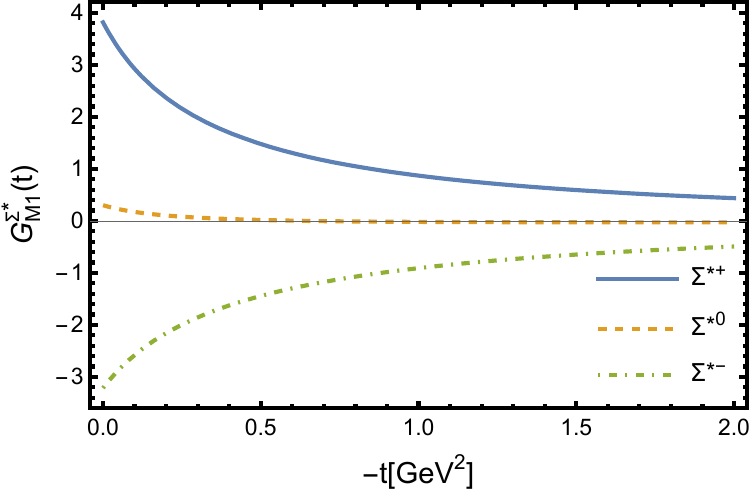}
	    \includegraphics[width=0.49\linewidth]{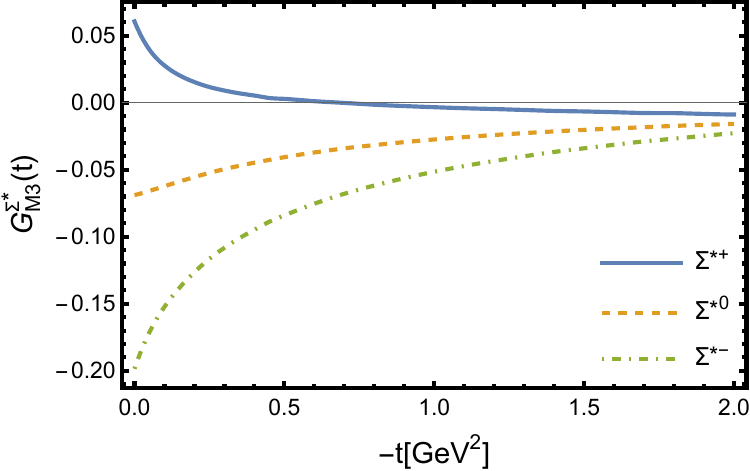}
    	\caption{\small{EMFFs of $\Sigma^*$. The solid, dashed and dot-dashed 
    	curves represent the EMFFs of $\Sigma^{*+}$, $\Sigma^{*0}$, 
    	and $\Sigma^{*-}$.}}
    	\label{sEM}
\end{figure}

\begin{figure}[htbp]
    	\centering
		\includegraphics[width=0.48\linewidth]{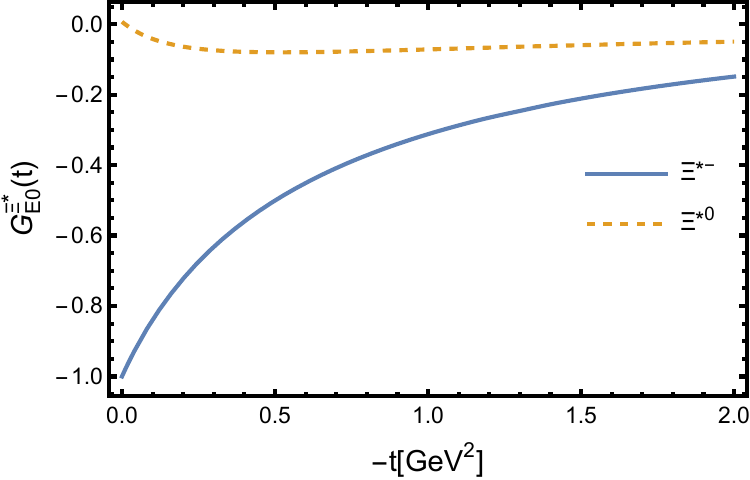}
	    \, \includegraphics[width=0.48\linewidth]{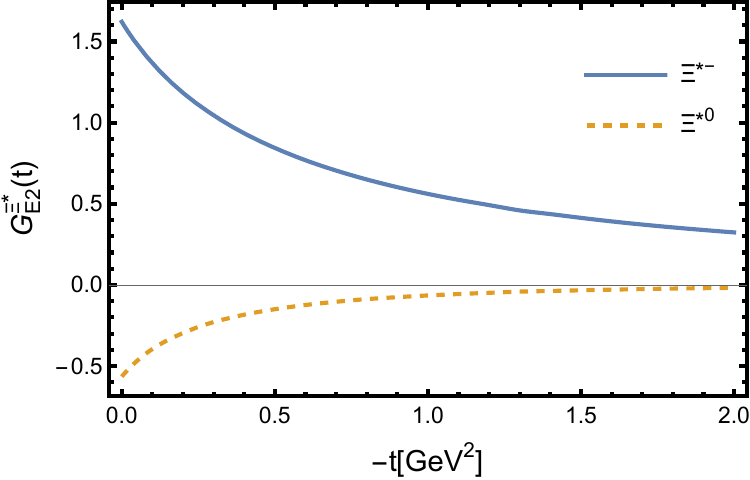}\\
		\, \includegraphics[width=0.47\linewidth]{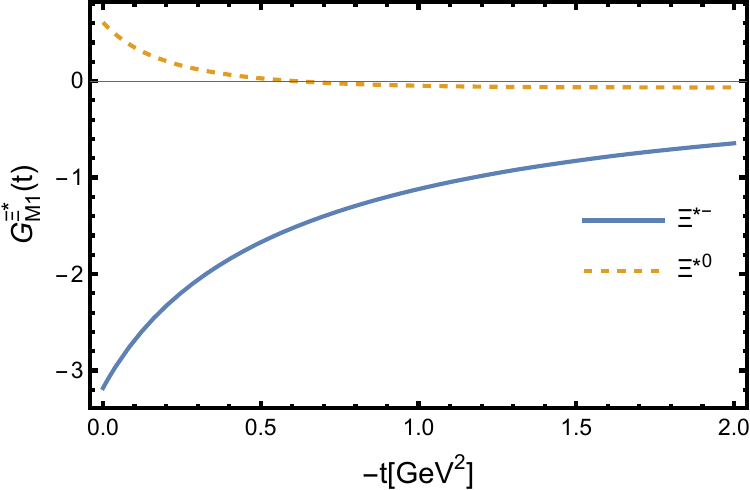}
	    \includegraphics[width=0.49\linewidth]{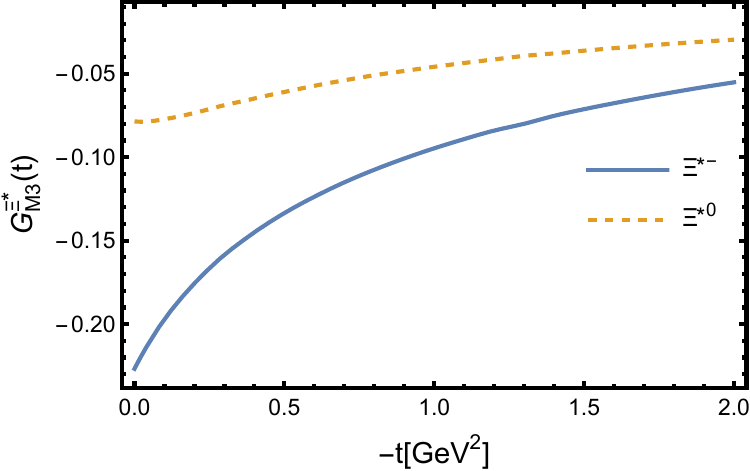}
    	\caption{\small{EMFFs of $\Xi^*$. The solid and dashed curves represent 
    	the EMFFs of $\Xi^{*-}$ and $\Xi^{*0}$.}}
    	\label{xEM}
\end{figure}

\begin{table}[htbp]
        \renewcommand\arraystretch{1.3}
		\centering
		\resizebox{\linewidth}{!}{ 
		\begin{tabular}{p{2.2cm} p{1.2cm}<{\centering} p{1.2cm}<{\centering} 
		p{1.2cm}<{\centering} p{1.2cm}<{\centering} p{1.2cm}<{\centering} 
		p{1.2cm}<{\centering} p{1.2cm}<{\centering} p{1.4cm}<{\centering} 
		p{1.2cm}<{\centering} p{1.2cm}<{\centering}}
     		\toprule
     		\toprule
     		${\langle r^2\rangle}_{E0}/ \text{fm}^2$ & $\Delta^{++}$ & 
     		$\Delta^{+}$ & $\Delta^{0}$ & $\Delta^{-}$ & $\Sigma^{*+}$ & 
     		$\Sigma^{*0}$  & $\Sigma^{*-}$ & $\Xi^{*0}$ & $\Xi^{*-}$ & 
     		$\Omega^-$\\
        	\midrule
        	This work & 0.894 & 0.894 & 0 & 0.894 & 0.784 & 0.087 & 0.614 
        	& 0.166 & 0.451 & 0.361\\
        	\midrule
        	LQCD~\cite{Alexandrou_2009,Alexandrou:2010jv} & $\cdots$ & 0.641(22) & 
            $\cdots$ & $\cdots$ & $\cdots$ & $\cdots$ & $\cdots$ & $\cdots$ & $\cdots$ & 0.355(14)\\
            LQCD~\cite{Boinepalli:2009sq} & $\cdots$ & 0.410(57) & 0 & $\cdots$ & 0.399(45) 
            & 0.020(7) & 0.360(32) & 0.043(10) & 0.330(20) & 0.307(15)\\
            $\chi$PT~\cite{Geng:2009ys} & 0.325(22) & 0.328(21) & 0.006(1) 
            & 0.316(23) & 0.315(21) & 0 & 0.315(21) & $-0.006(1)$ & 0.312(18) 
            & 0.307(15)\\
            $\chi$PT~\cite{Li:2016ezv} & 0.30(11) & 0.29(10) & $-0.02(1)$ 
            & 0.33(11) & 0.31(11) & 0 & 0.31(11) & 0.02(1) & 0.29(10) 
            & 0.27(10)\\
            $\chi$CQM~\cite{Berger:2004yi} & 0.43 & 0.43 & 0 & 0.43 & 0.42 
            & 0.37 & 0.03 & 0.06 & 0.33 & 0.29\\
            $\chi$QM~\cite{Wagner:2000ii} & 0.77 & 0.77 & 0 & 0.77 & 0.93 
            & 0.10 & 0.74 & 0.20 & 0.68 & 0.78\\
            1/$N_c$~\cite{Flores-Mendieta:2015wir} & 1.048 & 1.101 
            & 0.105 & 0.891 & 0.939 & $-0.031$ & 0.895 & $-0.098$ & 0.981 & 1.042\\
            1/$N_c$~\cite{Buchmann:2002et} & 0.783 & 0.783 & 0 & 0.783 
            & 0.869 & 0.108 & 0.669 & 0.206 & 0.561 & 0.457\\
            $\chi$QSM~\cite{Kim_Kim_2019} & 0.826 & 0.792 & $-0.069$ & 0.930 
            & 0.843 & $-0.024$ & 0.891 & 0.021 & 0.852 & 0.813\\
            
        	\bottomrule 
     	\end{tabular}
     	}
     	\caption{\small{Electric charge radii of the decuplet baryons, comparing with those from LQCD~\cite{Alexandrou_2009,Alexandrou:2010jv,Boinepalli:2009sq}, chiral perturbation theory~\cite{Geng:2009ys,Li:2016ezv}, and chiral constituent quark model~\cite{Berger:2004yi}, chiral quark model~\cite{Wagner:2000ii}, $1/N_c$ expansion~\cite{Buchmann:2002et,Flores-Mendieta:2015wir}, and chiral quark-soliton model~\cite{Kim_Kim_2019}.}}
     	\label{cr}
\end{table} 

    \begin{table}[htbp]
		\renewcommand\arraystretch{1.3}
		\centering
		\resizebox{\linewidth}{!}{ 
		\begin{tabular}{p{1.8cm} p{1.4cm}<{\centering} p{1.4cm}<{\centering} 
		p{1.4cm}<{\centering} p{1.4cm}<{\centering} p{1.4cm}<{\centering}	
		p{1.4cm}<{\centering} p{1.4cm}<{\centering} p{1.4cm}<{\centering} 
		p{1.4cm}<{\centering} p{1.4cm}<{\centering}}
     		\toprule
     		\toprule
     		${\langle r^2\rangle}_{M1}/ \text{fm}^2$ & $\Delta^{++}$ & 
     		$\Delta^{+}$ & $\Delta^{0}$ & $\Delta^{-}$ & $\Sigma^{*+}$ & 
     		$\Sigma^{*0}$ & $\Sigma^{*-}$ & $\Xi^{*0}$ & $\Xi^{*-}$ & 
     		$\Omega^-$\\
        	\midrule
        	This work & 0.827 & 0.827 & 0 & 0.827 & 0.703 & 0.421 & 0.573 
        	& 0.820& 0.443 & 0.340\\
        	\midrule
        	LQCD~\cite{Alexandrou:2010jv} & $\cdots$ & $\cdots$ & 
            $\cdots$ & $\cdots$ & $\cdots$ & $\cdots$ & $\cdots$ & $\cdots$ & $\cdots$ & 0.286(31)\\
        	$\chi$PT~\cite{Li:2016ezv} & 0.61(15) & 0.64(14) & 0.07(12) 
            & 0.55(19) & 0.59(16) & 0 & 0.59(16) & $-0.07(12)$ & 0.64(14) 
            & 0.70(12)\\
        	$\chi$QM~\cite{Wagner:2000ii} & 0.62 & 0.62 & 0 & 0.62 & 0.67 
        	& 0.82 & 0.61 & 0.82 & 0.58 & 0.53\\
            $\chi$QSM~\cite{Kim_Kim_2019} & 0.587 & 0.513 & 1.786 & 0.764 
            & 0.599 & 3.356 & 0.713 & 0.784 & 0.653 & 0.582\\
            
            \bottomrule 
        \end{tabular}
     	}
     	\caption{\small{Magnetic radii of the decuplet baryons, 
     	comparing with those from LQCD~\cite{Alexandrou:2010jv}, chiral perturbation theory~\cite{Li:2016ezv}, chiral quark model~\cite{Wagner:2000ii}, and chiral quark-soliton model~\cite{Kim_Kim_2019}.}}
     	\label{mr}
\end{table}
 
\begin{table}[htbp]
        \renewcommand\arraystretch{1.3}
		\centering
		\resizebox{\linewidth}{!}{ 
		\begin{tabular}{p{2.1cm} p{1.4cm}<{\centering} p{1.4cm}<{\centering} 
		p{1.4cm}<{\centering} p{1.4cm}<{\centering} 
		p{1.4cm}<{\centering} p{1.4cm}<{\centering} p{1.4cm}<{\centering} 
		p{1.4cm}<{\centering} p{1.4cm}<{\centering} p{1.4cm}<{\centering}}
     		\toprule
     		\toprule
     		$\mu/\mu_N$ & $\Delta^{++}$ & $\Delta^{+}$ & $\Delta^{0}$ & 
     		$\Delta^{-}$ & $\Sigma^{*+}$ & $\Sigma^{*0}$ & $\Sigma^{*-}$ & 
     		$\Xi^{*0}$ & $\Xi^{*-}$ & $\Omega^-$\\
        	\midrule
        	This work & 4.80 & 2.40 & 0 & $-2.40$ & 2.58 & 0.20 & $-2.18$ & 0.37 
        	& $-1.95$ & $-1.79$\\
        	\midrule
            
            PDG~\cite{ParticleDataGroup:2022pth} & 6.14(51) 
            & $2.7^{+1.0}_{-1.3} 
            \pm 1.5 \pm 3$ & $\cdots$ & $\cdots$ & $\cdots$ & $\cdots$ & $\cdots$ & $\cdots$ & $\cdots$ & $-2.02(5)$\\
            LQCD~\cite{Alexandrou_2009,Alexandrou:2010jv} & $\cdots$ & 1.91(16) & 
            $\cdots$ & $\cdots$ & $\cdots$ & $\cdots$ & $\cdots$ & $\cdots$ & $\cdots$ & $-1.835(94)$\\
            LQCD~\cite{Boinepalli:2009sq} & 3.20(56) & 1.60(28) & 0 
            & $-1.60(28)$ & 1.76(18) & 0.00(4) & $-1.75(13)$ & 0.08(5) 
            & $-1.76(8)$ & $-1.70(7)$\\
            LQCD~\cite{PhysRevD.79.051502} & 3.70(12) & 2.40(6) 
            & $\cdots$ & $-1.85(6)$ & $\cdots$ & $\cdots$ & $\cdots$ & $\cdots$ & $\cdots$ & $-1.93(8)$\\
            LQCD~\cite{Leinweber:1992hy} & 4.91(61) & 2.46(31) 
            & 0.00 & $-2.46(31)$ & 2.55(26) & 0.27(5) & $-2.02(18)$ & 0.46(7) 
            & $-1.68(12)$ & $-1.40(10)$\\
            LQCD~\cite{Lee:2005ds} & 5.24(18) & 0.97(8) & $-0.035(2)$ 
            & $-2.98(19)$ & 1.27(6) & 0.33(5) & $-1.88(4)$ & 0.16(4) 
            & $-0.62(1)$ & $\cdots$\\
            $\chi$PT~\cite{Geng:2009ys} & 6.04(13) & 2.84(2) & $-0.36(9)$ 
            & $-3.56(20)$ & 3.07(12) & 0 & $-3.07(12)$ & 0.36(9) & $-2.56(6)$ 
            & $-2.02$\\
            $\chi$PT~\cite{Li:2016ezv} & 4.97(89) & 2.60(50) & 0.02(12) 
            & $-2.48(32)$ & 1.76(38) & $-0.02(3)$ & $-1.85(38)$ & $-0.42(13)$ 
            & $-1.90(47)$ & $-2.02(5)$\\
            RQM~\cite{Schlumpf:1993rm} & 4.76 & 2.38 & 0 & $-2.38$ & 1.82 
            & $-0.27$ & $-2.36$ & $-0.60$ & $-2.41$ & $-2.35$\\
            QCDSR~\cite{Lee:1997jk} & 4.13(1.30) & 2.07(65) & 0 & $-2.07(65)$ 
            & 2.13(82) & $-0.32(15)$ & $-1.66(73)$ & $-0.69(29)$ & $-1.51(52)$ 
            & $-1.49(45)$\\
            $\chi$QM~\cite{Wagner:2000ii} & 6.93 & 3.47 & 0 & $-3.47$ 
            & 4.12 & 0.53 & $-3.06$ & 1.10 & $-2.61$ & $-2.13$\\
            $\chi$QSM~\cite{Kim_Kim_2019} & 3.65 & 1.72 & $-0.21$ & $-2.14$ 
            & 1.91 & $-0.04$ & $-1.99$ & 0.13 & $-1.84$ & $-1.69$\\
            HB$\chi$PT~\cite{Butler:1993ej} & 4.0(4) & 2.1(2) & $-0.17(4)$ 
            & $-2.25(25)$ & 2.0(2) & $-0.07(2)$ & $-2.2(2)$ & 0.10(4) & $-2.0(2)$ 
            & $-1.94(22)$\\
            1/$N_c$~\cite{Luty:1994ub} & 5.9(4) & 2.9(2) & $\cdots$ & $-2.9(2)$ 
            & 3.3(2) & 0.3(1) & $-2.8(3)$ & 0.65(20) & $-2.30(15)$ & $-1.94$\\
        	\bottomrule 
     	\end{tabular}
        }
        \caption{\small{Magnetic moments of the decuplet baryons, comparing with 
        those from PDG~\cite{ParticleDataGroup:2022pth}, 
        LQCD~\cite{Alexandrou_2009,Alexandrou:2010jv,Boinepalli:2009sq,Leinweber:1992hy,Lee:2005ds,PhysRevD.79.051502}, chiral perturbation theory~\cite{Geng:2009ys,Li:2016ezv}, relativistic quark model~\cite{Schlumpf:1993rm}, QCD sum rules~\cite{Lee:1997jk}, chiral quark model~\cite{Wagner:2000ii}, chiral quark-soliton mode~\cite{Kim_Kim_2019}, and $1/N_c$ expansion~\cite{Luty:1994ub}.}}\label{mm}
	\end{table}

    \begin{table}[htbp]
		\renewcommand\arraystretch{1.3}
		\centering
		\resizebox{\linewidth}{!}{ 
		\begin{tabular}{p{2.4cm} p{1.4cm}<{\centering} p{1.4cm}<{\centering} 
		p{1.4cm}<{\centering} p{1.4cm}<{\centering} p{1.4cm}<{\centering} 
		p{1.4cm}<{\centering} p{1.4cm}<{\centering} p{1.4cm}<{\centering} 
		p{1.4cm}<{\centering} p{1.4cm}<{\centering}}
     		\toprule
     		\toprule
     		$\mathcal{Q}/ \text{fm}^2$ & $\Delta^{++}$ & $\Delta^{+}$ &
     		$\Delta^{0}$ & $\Delta^{-}$ & $\Sigma^{*+}$ & $\Sigma^{*0}$ & 
     		$\Sigma^{*-}$ & $\Xi^{*0}$ & $\Xi^{*-}$ & $\Omega^-$\\
        	\midrule
            This work & $-0.075$ & $-0.037$ & 0 & 0.037 & $-0.045$ & $-0.006$ 
            & 0.033 & $-$0.009 & 0.027 & 0.024\\

            \midrule
            LQCD~\cite{Alexandrou_2009,Alexandrou:2010jv} & $\cdots$ & $-0.019(17)$ & 
            $\cdots$ & $\cdots$ & $\cdots$ & $\cdots$ & $\cdots$ & $\cdots$ & $\cdots$ & $0.019(3)$\\
            Skyrme~\cite{Oh:1995hn} & $-$0.088 & $-$0.029 & 0.029 & 0.088 
            & $-$0.071 & 0 & 0.071 & $-$0.046 & 0.046 & 0\\
            NQM~\cite{Krivoruchenko:1991pm} & $-$0.093 & $-$0.046 & 0 & 0.046 
            & $-$0.054 & $-$0.007 & 0.040 & $-$0.013 & 0.034 & 0.028\\
            QCDSR~\cite{Azizi:2009egn,Aliev:2009pd} & $-$0.028(8) & $-$0.014(4) 
            & 0 & 0.014(4) & $-$0.028(9) & 0.0012(4) & 0.03(1) & 0.0025(8) 
            & 0.045(15) & 0.12(4)\\
            $\chi$QM~\cite{Wagner:2000ii} & $-$0.252 & $-$0.126 & 0 & 0.126 
            & $-$0.123 & $-$0.021 & 0.082 & $-$0.030 & 0.048 & 0.026\\
            1/$N_c$~\cite{Buchmann:2002et} & $-$0.120 & $-$0.060 & 0 & 0.060 
            & $-$0.069 & 0.014 & 0.077 & $-$0.023 & 0.047 & 0.027\\
            GPM~\cite{Buchmann:2002xq} & $-$0.226 & $-$0.113 & 0 & 0.113 
            & $-$0.107 & $-$0.017 & 0.074 & $-$0.023 & 0.044 & 0.024\\
            $\chi$QSM~\cite{Kim_Kim_2019} & $-$0.102 & $-$0.039 & 0.023 
            & 0.085 & $-$0.070 & 0.003 & 0.077 & $-$0.016 & 0.069 & 0.061\\
            HB$\chi$PT~\cite{Butler:1993ej} & $-$0.08(5) & $-$0.03(2) 
            & 0.012(5) & 0.06(3) & $-$0.07(3) & $-$0.013(7) & 0.04(2) & $-$0.035(2) 
            & 0.02(1) & 0.009(5)\\
            
            \bottomrule
     	\end{tabular}
     	}
     	\caption{\small{Electric-quadrupole moments of the decuplet baryons comparing with those from LQCD~\cite{Alexandrou_2009,Alexandrou:2010jv}, Skyrme model~\cite{Oh:1995hn}, nonrelativistic quark model~\cite{Krivoruchenko:1991pm}, QCD sum rules~\cite{Azizi:2009egn,Aliev:2009pd}, chiral quark model~\cite{Wagner:2000ii}, $1/N_c$ expansion~\cite{Buchmann:2002et}, general QCD parameterization method~\cite{Buchmann:2002xq}, chiral quark-soliton model~\cite{Kim_Kim_2019}, and chiral perturbation theory~\cite{Butler:1993ej}.}}
     	\label{eqm}
	\end{table}

\subsection{GMFFs numerical results}    

\begin{figure}[htbp]
    	\centering
		\, \includegraphics[width=0.47\linewidth]{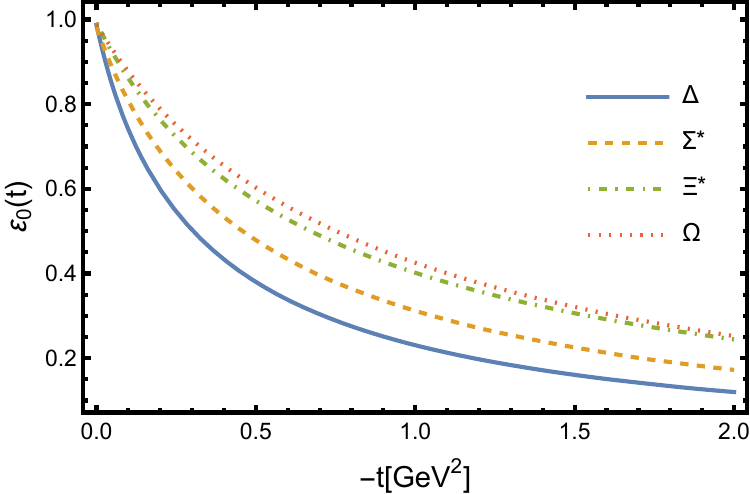}
	    \includegraphics[width=0.48\linewidth]{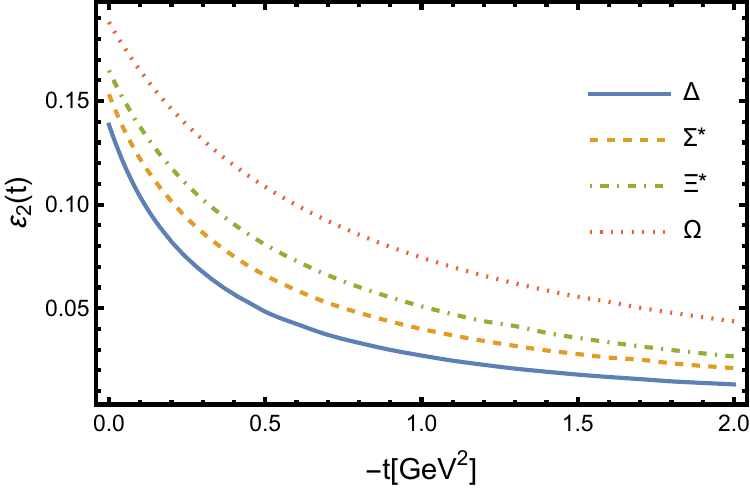}\\
		\,\, \includegraphics[width=0.48\linewidth]{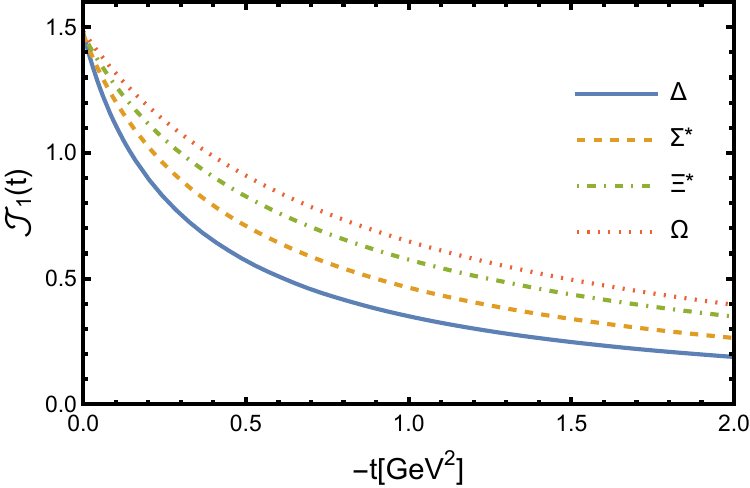}
	    \includegraphics[width=0.47\linewidth]{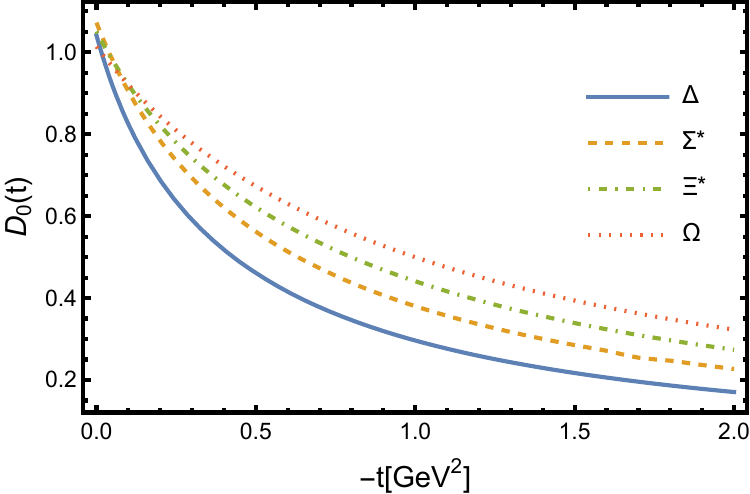}
    	\caption{\small{GMFFs of the decuplet baryons. The solid, dashed, 
    	dot-dashed, and dotted curves represent the GMFFs of $\Delta$, 
    	$\Sigma^*$, $\Xi^*$, and $\Omega$.}}
    	\label{GMFFs}
\end{figure}

\quad\, Figure~\ref{GMFFs} shows the obtained GMFFs including energy-monopole $\varepsilon_0(t)$, energy-quadrupole $\varepsilon_2(t)$, angular momentum-dipole $\mathcal{J}_1(t)$, and the D-term correlated $D_0(t)$. We employ the same normalization constant $\mathcal{C}$ and $\mathcal{C}_D$ with ones in EMFFs determined in Sec.~\ref{EMFFsnumericalresults}. 
It is seen that for all the decuplet baryons, $\varepsilon_0(0)$ and $\mathcal{J}(0)$ run 
from 0.97 to 0.99 and from 1.46 to 1.48 separately, which are almost consistent with the normalization 
condition $\varepsilon_0(0)=1$ and $\mathcal{J}(0)=3/2$. As discussed in Refs.~\cite{PhysRevD.78.094011,DAVIDSON1995163}, 
the momentum-dependent scalar function introduced in Eq.~\eqref{vertexfunction2} may break the gauge invariance 
and the electromagnetic Ward-Takahashi identity, and consequently the EMFFs and GFFs cannot be normalized at the 
same time. Similar with the discussion of the electric-quadrupole moment, the positive energy-quadrupole moment $\varepsilon_2(0)$ 
suggests that all the decuplet baryons have a prolate mass distribution. 

\begin{figure}[htbp]
    	\centering
		\includegraphics[width=0.47\linewidth]{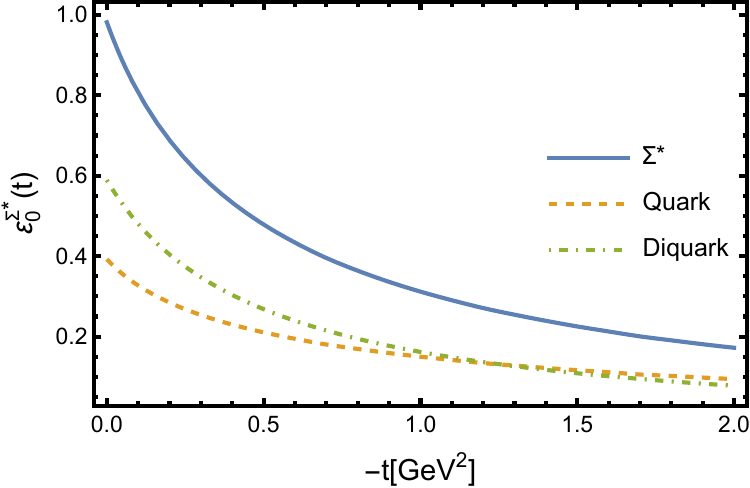}
	    \includegraphics[width=0.47\linewidth]{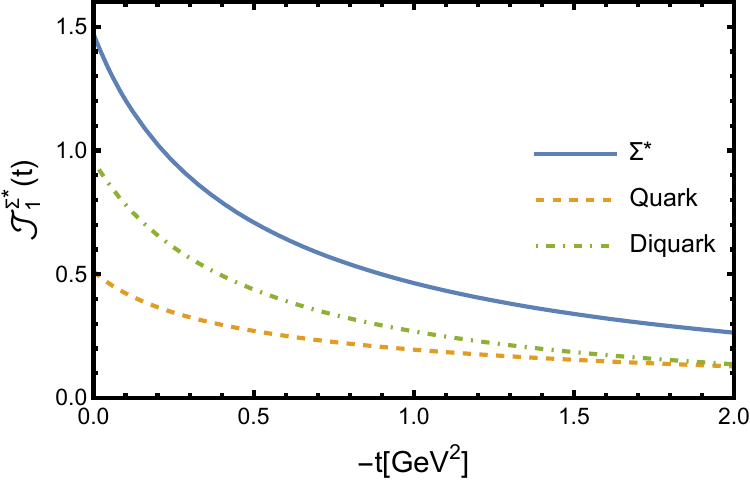}
    	\caption{\small{The energy-monopole and angular momentum form factors 
    	of $\Sigma^{*}$. The solid, dashed, and dot-dashed curves 
    	represent the total GMFFs and those contributed by quark 
    	and diquark.}}
    	\label{SGMFFs}
\end{figure}

Figure~\ref{SGMFFs} shows the energy-monopole and angular momentum-dipole form factors of $\Sigma^{*+}$ with the quark and the diquark contributions plotted respectively. According to Fig.~\ref{SGMFFs}, the angular-momentum contribution of the diquark is 
about twice that of the quark, especially when $t$ goes to 0. This phenomenon is consistent with our understanding on baryon spin since the decuplet baryons are composed of a spin-1/2 quark and a spin-1 diquark.

According to the definition in Eq.~\eqref{MassRadius}, we can further get the mass radii of the baryons as shown in 
Table~\ref{massr}. Comparing with the electric charge and magnetic radii in Tables~\ref{cr} and~\ref{mr}, the 
mass radii is a little smaller. Similarly, the mass radius becomes smaller as the mass increases.

\begin{table}[htbp]
		\centering
		\begin{tabular}{p{1.8cm} p{1.5cm}<{\centering} p{1.5cm}<{\centering} 
		p{1.5cm}<{\centering} p{1.5cm}<{\centering}}
     		\toprule
     		\toprule
       		${\langle r^2\rangle}_M/ \text{fm}^2$ & $\Delta$ & $\Sigma^*$ 
       		& $\Xi^*$ & $\Omega$\\
        	\midrule
        	This work & 0.801 & 0.516 & 0.368 & 0.298\\
        	\bottomrule
     	\end{tabular}
     	\caption{\small{Mass radii of the decuplet baryons.}}
     	\label{massr}
\end{table}    

The energy density, angular momentum density, and strong force density in the $r$-space inside the baryon can also be derived through the Fourier transformation. Refs.~\cite{Epelbaum:2022fjc,Diehl_2002,Freese_Miller_2022} 
suggest that the local density distribution must depend on the size of the wave packet of the system. An additional wave packet is necessary physically and mathematically to guarantee the convergence 
of the Fourier transformation. Of course, the wave package introduces a new parameter $\lambda$ and may have an influence on the 
definition of the radius~\cite{Epelbaum:2022fjc,Alharazin:2022xvp}. However, this issue is not a priority in this work.
    
Here we simply follow the idea of Refs.~\cite{Epelbaum:2022fjc,Diehl_2002,Freese_Miller_2022} 
and employ a Gaussian-like wave packet $e^{t/\lambda^2}$~\cite{Ishikawa:2017iym}. The parameter $\lambda$ 
has the mass dimension and $1/\lambda$ correlates with the size of the hadron. As seen in Table~\ref{massr}, the mass radii of the baryons become smaller as their mass increase. Ref.~\cite{Diehl_2002} has a detailed discussion on the determination of $\lambda$. 
For convenience and simplicity, we assume that $1/\lambda$ roughly relates to the Compton length of the system and there is a
linear relation between the mass radius and $1/\lambda$ of the baryon 
\begin{equation}
\label{lam}
    \sqrt{{\langle r^2\rangle}_M}=\alpha \frac{1}{\lambda},
\end{equation}
where the parameter $\alpha\sim 4$ is employed in our numerical calculation.

\begin{figure}[htbp]
    \centering
    \includegraphics[width=0.47\linewidth]{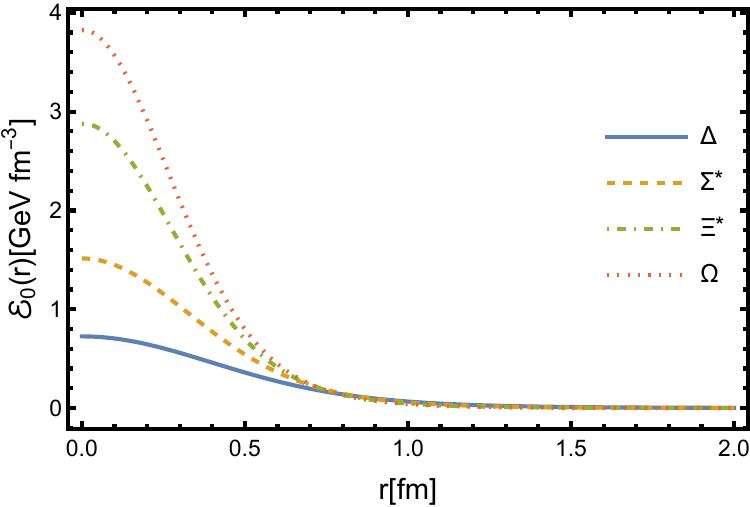}
	\includegraphics[width=0.47\linewidth]{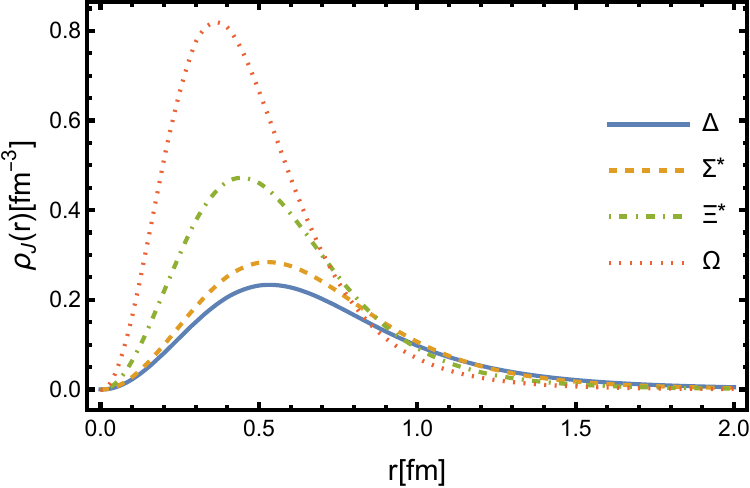}
    \caption{\small{The energy-monopole and angular momentum densities of 
    the decuplet baryons. The solid dashed, dot-dashed and dotted curves 
    represent the densities of $\Delta$, $\Sigma^*$, $\Xi^*$, and $\Omega$.}}
    \label{rep}
\end{figure} 

Figure~\ref{rep} shows the energy-monopole densities and angular momentum densities of the decuplet baryons.
Moreover, the integrated result of $\mathcal{E}_0(r)$ and $\rho_J(r)$ over the 
whole coordinate space gives the mass and spin of the corresponding baryon.
The right panel gives the angular momentum densities of the baryons and it is seen that the large $\lambda$ concentrates the densities 
close to the origin. 

Finally, $D_{0,2,3}(t)$ are supposed to connect with the pressure and shear force in the classical physical 
concept discussed in Ref.~\cite{Polyakov_2018}. As shown in Fig.~\ref{GMFFs}, the D-term, $D=D_0(0)$, of all the baryons are positive. However, 
it is argued that the D-term should be negative in order to guarantee the stability of the 
system in Ref.~\cite{PhysRevD.94.054024}. The sign of the present D-term is consistent with our previous 
results~\cite{PhysRevD.105.096002,Fu:2023ijy} in the same quark-diquark approach, and with the result of the hydrogen atom~\cite{Ji:2021mfb}. Although the phenomenon of 
$D >0$ does not consistent with the arguments in Ref.~\cite{PhysRevD.94.054024}, it still satisfies the von Laue condition 
$\int^ \infty _0 d r r^2 p_0(r)=0$. Here, we argue that the classical definitions of the pressure and shear force may not exist in the few-body system we are dealing with, 
because they are derived from the statistical means in the classical multi-body systems. The hydrogen atom is also a few-body system, so its non-positive D-term is not necessary. A more detailed discussion 
has been given in our work on $\Omega^-$ in Ref.~\cite{Fu:2023ijy}.

\section{SUMMARY AND DISCUSSION}\label{summary}
    
\quad\, In this work, the EMFFs and GFFs of all the decuplet baryons have been calculated 
systematically and simultaneously with a relativistic covariant quark-diquark approach. The baryons with spin-3/2 are considered as the combination of a quark and an axial-vector diquark and the total form factors are the sum of their 
contributions. To ensure the bound state between the quark and the diquark, an additional scalar function is used.
Although this scalar function may have an impact on the gauge invariance, the deviation of the normalization in our numerical results is small.

We then fit our results of the EMFFs to the LQCD calculations for $\Delta^+$ and $\Omega^-$ and try to find a set of 
parameters that give a systematical and reasonable description of all the 
decuplet baryons. Here, we simply keep the parameters for the $s$ quark system of $\Omega^-$ and re-determine the others containing $u$ and $d$ quarks.
The model parameters cannot be rigorously determined due to the lack of experimental and LQCD data on the strongly parameter-dependent higher-order multipole form factors.

In the numerical calculations, we obtain the electromagnetic properties including electric charge radii, magnetic moments, 
electric-quadrupole moments, and magnetic octupole moments, which are in a reasonable agreement with those from some experiments, LQCD calculations, and other models.
Moreover, we also calculate the GMFFs of the decuplet baryons, and derive the mechanical properties of the systems, such as their mass radii, energy and angular momentum distributions. It shows that the mass radius is smaller than the electromagnetic radius for all the baryons, and the mass radii become small as the baryon mass increase.
Moreover, the distributions in the coordinate space for the energy and angular momentum distributions are also shown with introducing an effective wave package. The sign of the D-terms in our approach remains positive and how to understand it is still controversial.

It is expected that the present systematical description of the EMFFs and GFFs for all the decuplet 
baryons might provide more useful information to comprehend the inner structure of those baryons with spin-3/2, and also provide 
reference for future possible experiments at EIC, EicC, and JPARC.

\section*{ACKNOWLEDGEMENTS}
\quad\, This work is supported by the National Key Research and Development Program of China under Contracts No. 2020YFA0406300 
and the National Natural Science Foundation of China under Grants No. 11975245 and No. 12375142. it is also supported by 
the Sino-German CRC 110 “Symmetries and the Emergence of Structure in QCD” project by NSFC under Grant No. 12070131001 
and the Key Research Program of Frontier Sciences, CAS, under Grant No. Y7292610K1.

\appendix

\renewcommand\thesection{Appendix~\Alph{section}}
\section{SU(6) WAVE FUNCTIONS OF THE DECUPLETS}\label{appendix}
\par\noindent\par\setcounter{equation}{0}
\renewcommand{\theequation}{A\arabic{equation}}

\quad\, In the quark-diquark approach, the spin-isospin SU(6) wave 
functions of the decuplets are expressed as~\cite{Lichtenberg_Tassie_Keleman_1968}

\begin{align}
    &\ket{\Delta^{++}} =\ket{u(uu)}\phi, \\
    &\ket{\Delta^{+}} 
    =\left(\sqrt{\frac{2}{3}}\ket{u(ud)}+\sqrt{\frac{1}{3}}\ket{d(uu)}
    \right)\phi, \\
    &\ket{\Delta^{0}} 
    =\left(\sqrt{\frac{2}{3}}\ket{d(ud)}+\sqrt{\frac{1}{3}}\ket{u(dd)}
    \right)\phi, \\
    &\ket{\Delta^{-}} =\ket{d(dd)}\phi ,\\
    &\ket{\Sigma^{*+}} 
    =\left(\sqrt{\frac{2}{3}}\ket{u(us)}+\sqrt{\frac{1}{3}}\ket{s(uu)}
    \right)\phi, \\
    &\ket{\Sigma^{*0}} 
    =\left(\sqrt{\frac{1}{3}}\ket{d(us)}+\sqrt{\frac{1}{3}}\ket{u(ds)}
    +\sqrt{\frac{1}{3}}\ket{s(ud)}\right)\phi, \\
    &\ket{\Sigma^{*-}} 
    =\left(\sqrt{\frac{2}{3}}\ket{d(ds)}+\sqrt{\frac{1}{3}}\ket{s(dd)}
    \right)\phi, \\
    &\ket{\Xi^{*0}} =\left(\sqrt{\frac{2}{3}}\ket{s(us)}
    +\sqrt{\frac{1}{3}}\ket{u(ss)}\right)\phi, \\
    &\ket{\Xi^{*-}} =\left(\sqrt{\frac{2}{3}}\ket{s(ds)}
    +\sqrt{\frac{1}{3}}\ket{d(ss)}\right)\phi, \\
    &\ket{\Omega^{-}} =\ket{s(ss)}\phi,
\end{align}
where $\phi$ is the spin wave function and $(q_a q_b)$ stands for the axial-vector diquark which is composed of quarks $q_a$ and $q_b$.

\bibliographystyle{unsrt}
\bibliography{references}

\end{document}